# Temporal Plasmonics: Fano and Rabi regimes in the time domain in metal nanostructures

Oscar Ávalos-Ovando,[*1,2] Lucas V. Besteiro,[1,3] Zhiming Wang,[*1] and Alexander O. Govorov[*1,2]

[1] Institute of Fundamental and Frontier Sciences, University of Electronic Science and Technology of China, Chengdu 610054, China

[2] Department of Physics and Astronomy, Ohio University, Athens, Ohio 45701, United States

[3] Centre Énergie Matériaux et Télécommunications, Institut National de la Recherche Scientifique, 1650 Boul. Lionel Boulet, Varennes, QC J3X 1S2, Canada

*oa237913@ohio.edu, jmwahng@gmail.com, govorov@ohio.edu

**Abstract:** The Fano and Rabi models represent remarkably common effects in optics. Here we study the coherent time dynamics of plasmonic systems exhibiting Fano and Rabi spectral responses. We demonstrate that these systems show fundamentally different dynamics. A plasmonic system with a Fano resonance displays at most one temporal beat under pulsed excitation, whereas a plasmonic system in the Rabi-like regime may have any number of beats. Remarkably, the Fano-like systems show time dynamics with very characteristic coherent tails despite the strong decoherence that is intrinsic for such systems. The coherent Fano and Rabi dynamics that we predicted can be observed in plasmonic nanocrystal dimers in time-resolved experiments. Our study demonstrates that such coherent temporal plasmonics includes nontrivial and characteristic relaxation behaviors and presents an interesting direction to develop with further research.

## 1 Introduction

The Fano and Rabi effects are characteristic properties of two interacting oscillators. For a system to exhibit the Fano effect (**FE**), one oscillator should have a narrow absorption resonance and the other one should be strongly damped [1]. In striking contrast to the Fano system, the two oscillators involved in a Rabi resonance have narrow absorption lines. Importantly, in both models, the involved oscillators should interact, and this interaction leads to interesting consequences. In the Fano system, the optical resonance possesses a peculiar lineshape, whereas the Rabi system exhibits a very characteristic splitting in its optical spectrum.

Modern nanostructures offer a variety of optical systems where the FE can be realized [2, 3]. In the field of nanostructures, prominent examples of FE come from exciton-plasmon interaction, plasmon-plasmon coupling and the interaction of a localized exciton with a continuous spectrum [4-18]. Typically, the FE is observed in the spectral responses, such as absorption or scattering, which are optical measurements in the frequency domain. In contrast, this paper focuses on manifestations of the FE in the time domain under pulsed excitations. We will contrast the time-resolved FE with the case of Rabi resonance in the time domain.

Regarding the Rabi regime, this effect has been observed in many experimental systems in both spectral and time-resolved domains. In hybrid exciton-plasmon nanostructures and in electromagnetic resonators with a built-in two level system, this phenomenon is detected as the so-called Rabi splitting in the frequency domain [19-22]. For plasmonic nanocrystals, this spectral splitting effect is often regarded as a plasmon-plasmon hybridization [23]. In quantum semiconductor systems, it can be seen as the so-called Rabi oscillations in the time domain, when a two-level excitonic system interacts with a monochromatic electromagnetic wave [24-26]. In all the above cases, the Rabi-like regime involves two different excitations, both having narrow absorption lines.

This study investigates the Fano and Rabi regimes in the time domain using coupled plasmonic nanoparticles (NPs) as a model system. We found that the FE in the time domain exhibits very characteristic and unique features. The dynamical response of a Fano system is qualitatively different from that of a Rabi resonance. In the case of the Fano resonance, the coherent relaxation exhibits at most one beat in the dynamics and involves two relaxation exponents, fast and slow. In the Rabi-like systems, the relaxation dynamics may have a large number of beating oscillations before they are fully dampened. To reveal these features of coherent relaxation, we employed two different excitation methods: point dipoles placed next to the coupled NPs and external electromagnetic pulses. The models used in our study include NPs made of both Drude and real noble metals. Our study aims to motivate a novel direction of research termed by us as "coherent temporal plasmonics", in which custom-made plasmonic nanostructures could provide a very convenient platform for observing complex and nontrivial dynamical regimes.

## 2  Formalism

The formalism used in this study is based on the Maxwell's equations in the time domain and the local dielectric function model. We use Comsol Multiphysics software that specializes in the frequency-domain calculations, and, therefore, it will be convenient for us to utilize the Fourier transform. The displacement field, $\mathbf{D}$, in our formalism is given by (see Supporting Information)

$$\mathbf{D}(\mathbf{r},t) = \int_{-\infty}^{\infty} e^{+i\omega t} \mathbf{D}_\omega(\mathbf{r}) \cdot d\omega = \int_{-\infty}^{\infty} e^{+i\omega t} \varepsilon_0 \varepsilon_\omega(\mathbf{r}) \cdot \mathbf{E}_\omega(\mathbf{r}) \cdot d\omega , \qquad (1)$$

where $\mathbf{D}_\omega(\mathbf{r})$ and $\mathbf{E}_\omega(\mathbf{r})$ are the Fourier transforms of the displacement and electric fields, respectively; $\varepsilon_\omega(\mathbf{r})$ is the local dielectric constant in the frequency domain and it should obey the property: $\mathrm{Im}\,\varepsilon_\omega < 0$ for $\omega > 0$. More details for this formalism are given in the Supporting Information.

Figure 1 shows the typical physical settings used in our study. To probe the coherent dynamics of coupled NPs, we either place an excitation dipole next to the NP pair (Fig. 1a) or use an incident electromagnetic plane wave (Fig. 1d). Assuming a pulsed excitation, the temporal responses of our system can be conveniently written via the corresponding Green functions, $G_E(\omega)$ and $G_d(\omega)$:

$$d_{NC,E}(t) = \int_{-\infty}^{\infty} e^{+i\omega t} E_0(\omega) G_E(\omega) d\omega , \qquad d_{NC,d}(t) = \int_{-\infty}^{\infty} e^{+i\omega t} d_0(\omega) G_d(\omega) d\omega , \qquad (2)$$

where $E_0(\omega)$ and $d_0(\omega)$ are the Fourier transforms from the external electric field (x-component) and the exciting dipole, respectively. Then, the total dipole moment in a NP dimer is defined as $d_{tot} = d_{NP1} + d_{NP2}$. For both types of excitation, we used short Gaussian pulses. In the case of the exciting dipole, the pulse function is given by $d_0(t) = d_0 \exp[i\omega_0 t - (t/\Delta t)^2]$, where $d_0$, $\omega_0$ and $\Delta t$ are the pulse parameters: the amplitude, the central frequency and the pulse duration, respectively.

## 3 Results

We now begin constructing our main model systems (Rabi- and Fano-like) by incorporating NPs made of "Drude" metals (Fig. 1). These cases allow us to understand general properties of the Fano and Rabi dynamics without dealing with the complications of real metals, such as interband transitions. Therefore, we take the dielectric constant of our NPs in the standard form (see Supporting Information and [27]):

$$\varepsilon_i = \varepsilon_{i,b} - \frac{\omega_{i,p}^2}{\omega \cdot (\omega - i\gamma_i)}, \qquad (3)$$

where $i$ is the index running over the NP labels. The medium outside the NPs is water (i.e. $\varepsilon_w = 1.8$). We model two Drude dimers involving two dielectric functions, Drude₁ and Drude₂, which define the sharp and broad resonances, respectively (Fig. 1). The corresponding Drude parameters can be found in Supporting Information. The Rabi dimer includes two NPs with the same dielectric constant Drude₁. The Fano dimer in Fig. 1d comprises two nanorods with different dielectric functions. Although these cases are only a physical illustration, we found that the results obtained for the time dynamics are very generic. For the Rabi dimer, we first calculated the extinction cross section for one component (Fig. 1b, left), and we observed only one sharp dipolar resonance at 481 nm. Then, when the dimer system is excited with a point dipole pulse $d_0(t)$, we see that the absorption spectra exhibit multiple peaks. The two main resonances around 475 and 508 nm (Fig. 1b, right) correspond to the strongest anti-symmetric and symmetric modes, denoted as A- and S-modes (Fig. S2). The time dynamics for the total moment is shown in Fig. 1c. This trace shows a sustained oscillation with time, with several beats and nearly no decay in the time range shown, long after the initial pulse duration (shown in blue). Longer times start to show the decay. The frequency of the beats corresponds well to the difference between the frequencies of the main modes in the dimer, $\omega_{\text{A-mode}} - \omega_{\text{S-mode}}$, in Fig. 1b.

Regarding the Fano dimer, NP₂ shows a broad resonance in the extinction owing to its dielectric constant, Drude₂, while NP₁ exhibits a sharp peak (Fig. 1e). In the dimer configuration, these two, very different resonances will interfere destructively and give rise to a Fano resonance lineshape in the extinction, as shown in Fig. 1f (and Fig. S1b). Let's now look at the time evolution of the Fano dimer upon excitation with a 3-fs electromagnetic pulse. We observe now an interestingly different response: no beats and a sustained oscillation of the dimer's total dipole moment at long times. The slow exponential decay for the signal's envelope with an exponent of $\tau_s \sim 45$ fs persists for long after the initial pulse has ended, as shown in Fig. 1f. This slow exponent corresponds to the damping of an excitation that is mostly localized in NP₁. However, NP₁ in our dimer is strongly interacting with NP₂, and the exponent $\tau_s$ is significantly reduced from the original decay in the $NP_1$, $2/\gamma_{NP_1}$. At short times just after the pulse, we can also see another, faster relaxation, with an exponent of $\tau_f \sim 3\,\text{fs}$. This exponent comes from the fast decay inside the strongly dissipative $NP_2$. The single maximum in the amplitude comes from the confluence of two reasons: the initial pulse and the fast decay dynamics in $NP_2$. As we will soon see with more examples below, the whole pattern of relaxation in Fig.1f is very characteristic for the physical Fano system.

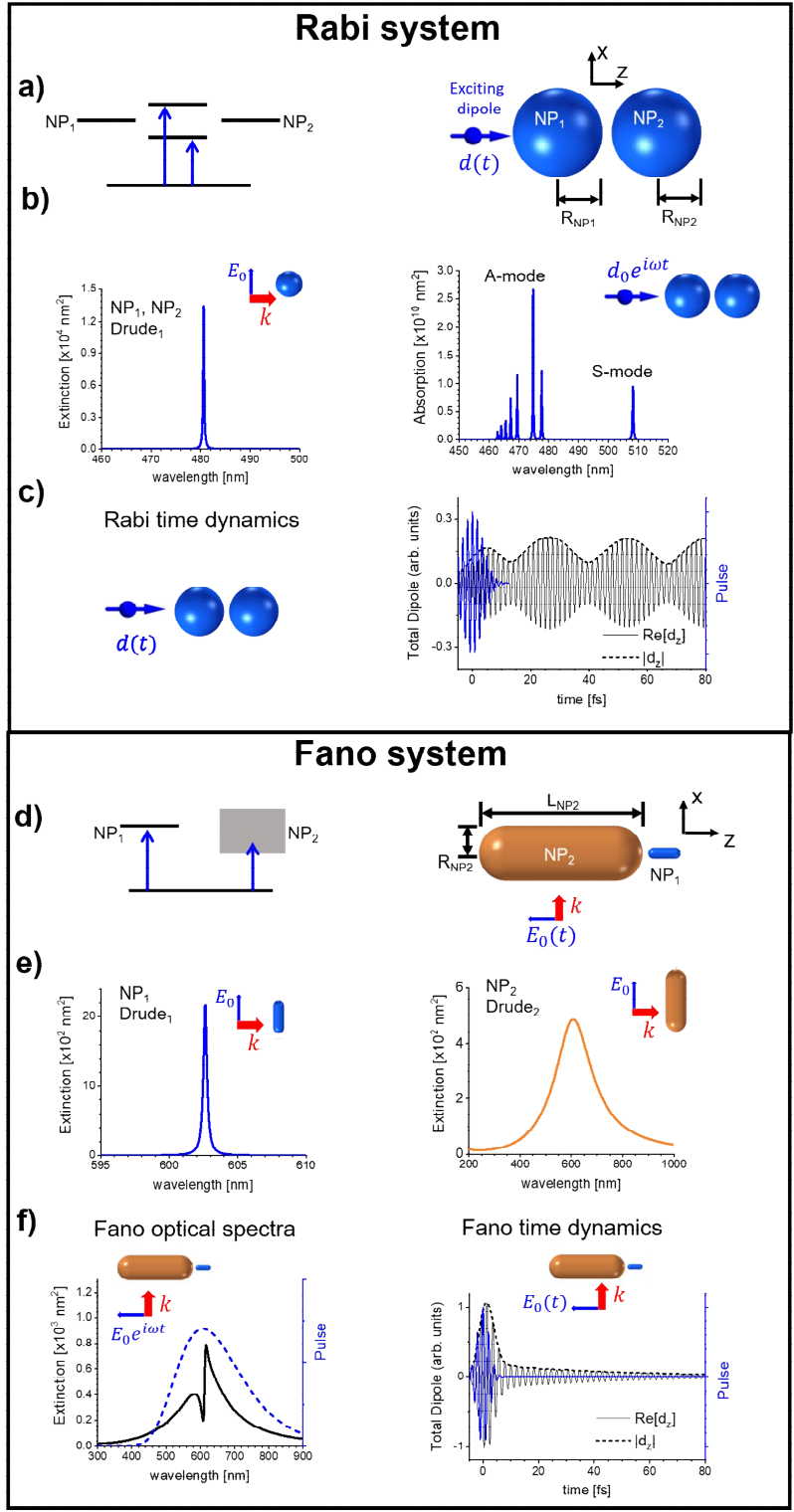

**Fig. 1:** Schematic and properties of Drude-Rabi (a,b,c) and Drude-Fano (d,e,f) systems. (a) Energy level scheme (left) and the geometry of the Rabi dimer under dipole excitation (right). (b) Extinction of one isolated NP (left) and the absorption spectrum of the dimer (right). (c) Dimer's time dynamics of the total dipole moment, including the excitation pulse (blue). Similarly, in the following panels, we show the optical properties of the Fano dimer. (d) Energy diagram of the Fano dimer. (e) Extinctions of isolated NPs. (f) Dimer's extinctions and time dynamics. For the Rabi system, we used $R_{NP1}=R_{NP2}=5$ nm, gap=1 nm, $\Delta t=5$ fs, $\lambda_0=495$ nm. For the Fano system, we took $R_{NP1}=1$ nm, $L_{NP1}=6$ nm, $R_{NP2}=5$ nm, $L_{NP2}=30$ nm, gap=1 nm, $\Delta t=3$ fs, and $\lambda_0=608$ nm, where $\lambda_0 = 2\pi c / \omega_0$. For the linearly polarized light, we assume a wave polarized along z, incoming with k∥x.

The above model systems provided us with interesting physical behaviors, and now it is time to look at these two regimes in real materials. In order to show that the properties seen in the Drude models are general, we now turn to model realistic materials such as gold and silver for the NPs [28], studying how the interacting plasmonic resonances in real dimers will interfere and create non-trivial regimes in their time dynamics. First, we characterize isolated single NPs. Ag-NPs show typically sharp plasmonic resonances and long decay times in their time dynamics for the total dipole moment, whereas Au-NPs show wider resonances and shorter decay times (Fig. S3). These single NPs are now combined to build a Ag-Ag Rabi dimer (Fig. 2), and a Au-Ag Fano dimer (Fig. 3).

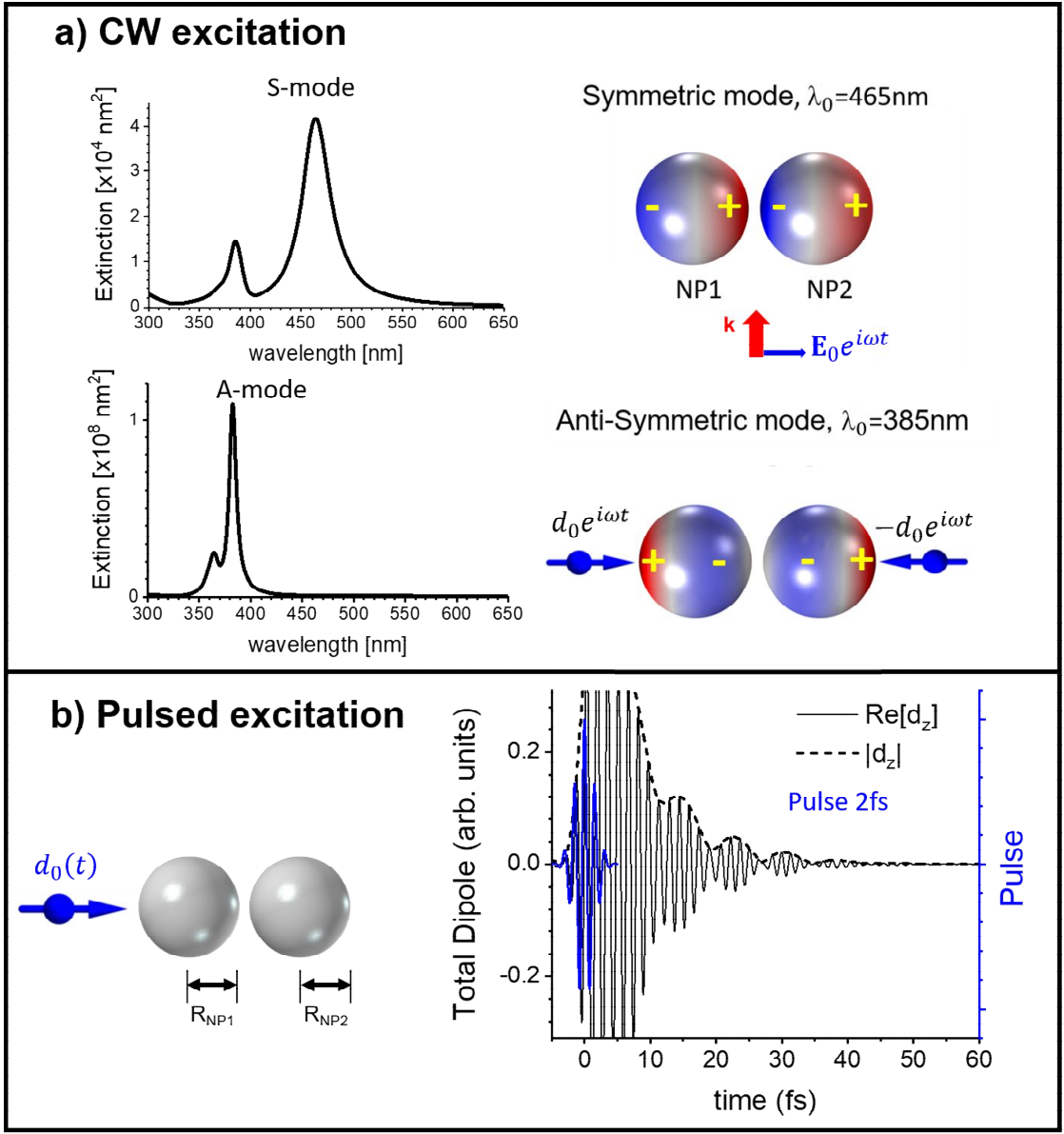

**Fig. 2:** (a) Extinction spectra and surface charge density maps for the Ag-Ag Rabi dimer under CW excitation. We use here two approaches: the electromagnetic wave excitation and the excitation with two anti-symmetric dipoles. In this way, we can see both symmetric and anti-symmetric modes in our Rabi dimer. (b) Dimer's time dynamics under the dipole excitation for a pulse at $\lambda_0$=465 nm. We used $R_{NP1}$=$R_{NP2}$=20 nm, gap=4 nm, and a pulse of $\Delta t$=2 fs (shown in blue).

Figure 2 shows the Ag-Ag Rabi dimer extinction cross sections under continuous wave (CW) with either a plane wave or with two point dipoles, and the time dynamics under the pulsed excitation. In Fig. 2a, the Rabi dimer exhibits the S- and A-modes as plasmonic peaks at 465 nm and 385 nm, respectively. Those peaks were revealed using 2 fs pulses. In Fig. 2b, the time dynamics of the total dipole moment show that $d_{tot}$ oscillates and decays after the initial excitation, showing up to 4 significant beats for $1\,\text{fs} < t < 40\,\text{fs}$. This means that the plasmon energy becomes transferred back and forth between the NPs, for much longer times than the duration of the initial pulse, eventually decaying. We see that this case is analogous to the Rabi dimer made of the Drude metal (Fig. 1c) but, unlike the multiple beats with almost no decay in the Drude case, the Rabi dimer made of a real metal results in a much faster decay, although still showing several beats. This multi-beat behavior is certainly the key feature of the time dynamics of the Rabi-resonance systems. Again, the frequency of the beats in the dynamics in Fig. 2b is given by the difference between the frequencies of the modes, $\omega_{\text{A-mode}} - \omega_{\text{S-mode}}$, since these modes (A- and S-modes) dominate the response of the dimer in the frequency domain (Fig. 2a).

In the next step, we look at the results for the Au-Ag Fano dimer (Fig. 3). Unlike the Rabi bi-harmonic mode, this Fano dimer is multi-harmonic, which is evidenced as a broad spectrum with the destructive interference dip in the extinction spectra at 591 nm (Fano antiresonance), surrounded by two maxima, at 554 nm and 618 nm. Dipoles in each NP are schematically shown as black arrows for each resonance. Then, the dimer is excited with a pulsed point dipole, as schematically depicted in Fig. 3b, using a 3-fs pulse centered at the Fano antiresonance (Fig. 3a). The oscillation of the total dipole, $d_{tot}$, shows a single pronounced beat after the pulse, a fundamental difference with respect to the Rabi dimer, which typically shows several. We note that we do not consider as a beat the first maximum of the envelope function in Fig. 3b, since it reflects the excitation pulse. At long times, we see again the decay with the long characteristic tail falling as $\exp[-t / \tau_s]$, like for the Drude case in Fig. 1f. In this case with the Au-Ag dimer, we have for the slow decay time: $\tau_l \sim 10\,\text{fs}$; this number reflects the plasmonic decay inside the metals. The dynamical behavior in Fig. 3b is the signature of the Fano coupling in the time domain. A coupling of this kind exhibits only one beat and leads to the long-time coherent energy dissipation within the Fano dimer.

To summarize our observations regarding the Fano effect in the time and frequency domains, we look now at several related parameters. The most interesting observation is that Fano dynamics have at most only one beat, whereas Rabi dynamics may have virtually any number of beats. In Fig. 4a (left), we show the data for an interval limited to 80 fs. The Rabi dynamics for the Drude and Ag NPs have 4 and 7 beats in that interval, respectively (see Figs. 1c and 2b). The Au-Ag Fano dimer exhibits only two temporal maxima in the amplitude, from which only one is a beat (Fig. 3b). In the Fano dynamics of the Au-Ag dimer, the first temporal maximum comes simply from the excitation pulse and the second one appears as a beat, due to the coherent energy transfer between NPs (Fig. 3b). For the Fano dimers composed of Drude NPs, we do not see beats, however; simultaneously, the dynamic trace contains a very characteristic tail, which is typical for the Fano dynamics (Fig. 1f and Fig. S12). This fundamental difference between the Rabi and Fano cases comes from the fact that the response of a Rabi system in the frequency domain is bi-harmonic (Fig. 1b), whereas the spectrum of a Fano system in the frequency domain is multi-harmonic (Fig. 1f). Physically, Rabi dynamics showcased here should be more coherent since it arises in two oscillators with weak decoherence. However, the Fano system should be less coherent since it has one oscillator with very fast decay. Indeed, we observe such property. Remarkably, despite the strong decoherence in the Fano systems, we actually observe their characteristic and prominent coherent dynamics, in the form of the single beat and the slowly decaying tail at long times. The fundamental differences between the dynamical responses of the considered systems can be also seen from the other panels in Fig. 4. Here we need to look at the following parameters: $R_\tau = \tau_{rel,NP2} / \tau_{rel,NP1}$, $R_{OS} = OS_{NP2} / OS_{NP1}$ and $R_\gamma = \gamma_{NP2} / \gamma_{NP1}$, where , $\tau_{rel,i}$, $OS_i$, and $\gamma_i$ are the relaxation time, the oscillator strength and the plasmon resonance broadening, respectively, for the $i$-th NP. In the Rabi model, the above ratios are all around unity, and this property leads to the typical bi-harmonic behavior. In contrast to the Rabi model, the Fano dynamics exhibit fundamentally different spectral properties: $R_\tau \ll 1$, $R_{OS} \gg 1$ and $R_\gamma \gg 1$ (Fig. 4). The above inequalities are the characteristic properties of a Fano system, which should contain one oscillator with a broad resonance and one with a narrow absorption line. The last parameter to consider in Fig. 4b is the ratio of the peak extinction cross sections, $R_{peak} = \sigma_{peak,NP2} / \sigma_{peak,NP1}$. Interestingly, $R_{peak}$ can be both small and large in the Fano case. Therefore, the latter parameter cannot be used to discriminate between the Fano and Rabi regimes.

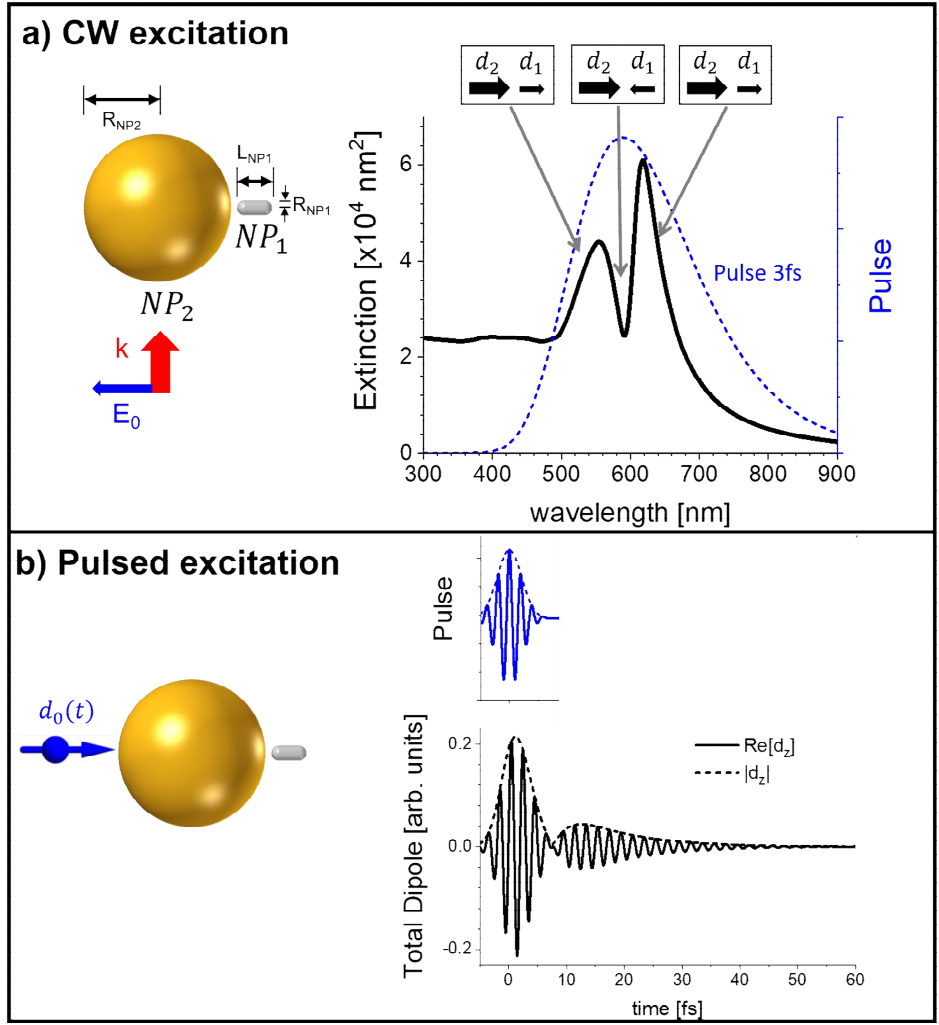

**Fig. 3:** Dynamic properties of the Au-Ag Fano dimer. (a) Extinction of the dimer for the electromagnetic CW excitation. Here we also show the structure of the induced dipoles in the dimer, which leads to the Fano effect. (b) Dimer's time dynamics under the dipole excitation for a pulse with $\lambda_0$=589 nm. We used $R_{Au}$=50 nm, $R_{Ag}$=5 nm, $L_{Ag}$=27 nm, gap=4 nm, and a pulse of $\Delta t$=3 fs. The spectrum for the dipole excitation pulse is shown in blue.

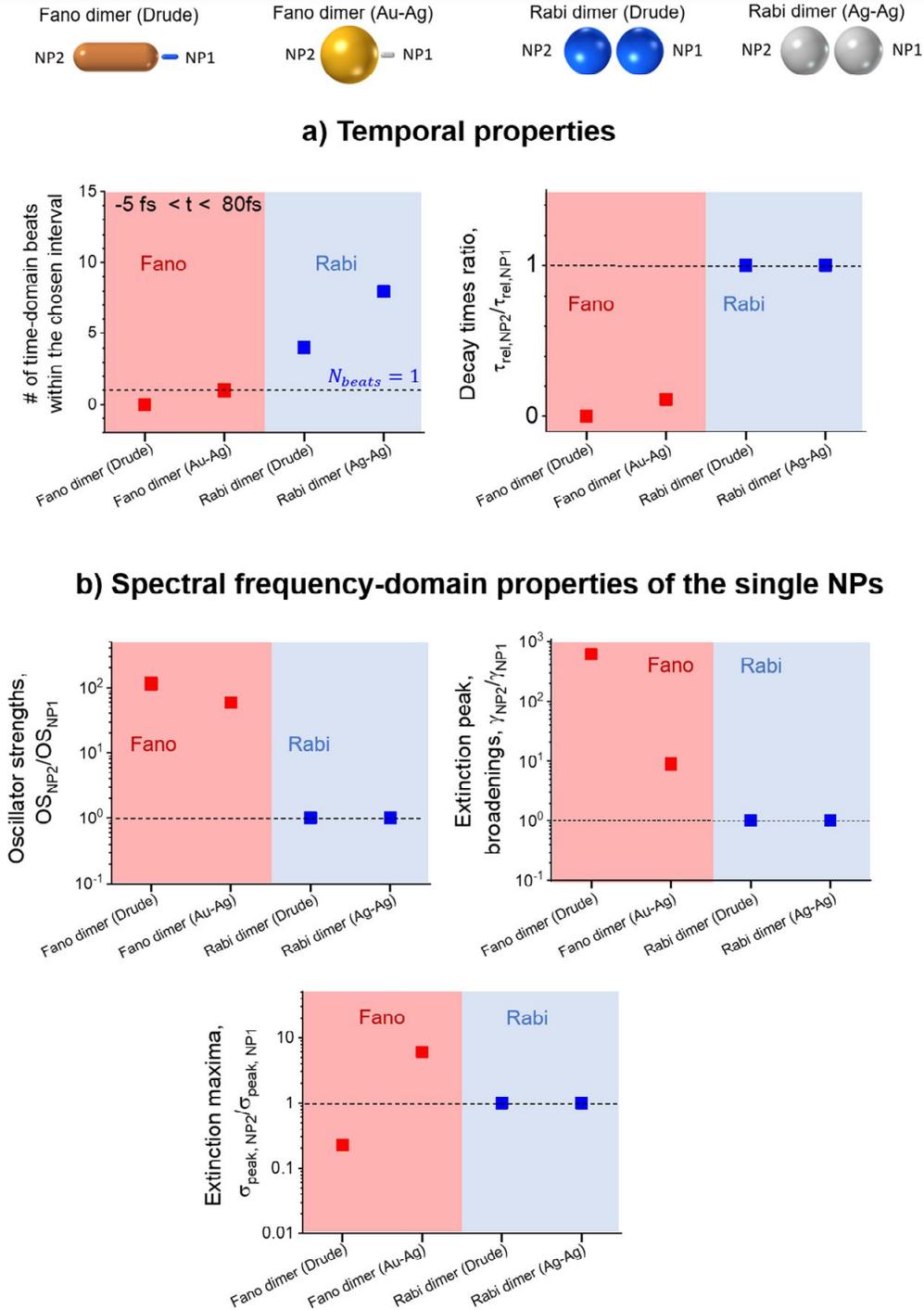

**Fig. 4:** Summary with several kinetic parameters, for the Fano (red) and Rabi (blue) regimes, for all the dimers presented in this study (schematically shown in the upper row). a) Temporal properties of the dimer's numbers of beats (left panel), and of the decay-time ratios of isolated NPs (right panel). b) Ratios of properties of single isolated NPs in the frequency domain, for the oscillator strength (top-left panel), the plasmon extinction broadenings (top-right panel) and the extinction maxima (bottom panel).

Another property that we observed in our calculations is non-locality of the Fano dynamics. Different responses from the same Fano dimer may show qualitatively different time traces. For example, the coherent beat for the Au-Ag Fano dimer does not appear if we excite the system with a point dipole placed next to the Ag NP (Fig. S9). This property also concerns local electric fields in a NP complex. In Fig. S10 for the Fano dimer made of Drude metals, we see that the dynamic electric field displays the characteristic coherent beat only inside the $NP_2$, whereas the fields between the NPs and inside the smaller NP do not show such a feature.

In our study, the Drude NPs serve as convenient physical models, whereas Au and Ag NPs in a liquid matrix represent realistic experimental systems. Our choice of shapes, sizes and materials in the Fano and Rabi dimers is based on the following considerations: (1) The Rabi dimer is designed with the Ag NPs since silver NPs exhibit a narrow and strong plasmonic resonance. In this case, the resonant interaction of two Ag NPs can be strong, and the resulting collective spectrum can show the Rabi-splitting effect. Indeed, we see this splitting effect in the computed spectra in Fig. 2. Furthermore, to obtain a prominent Rabi splitting, the distance between the NPs should be relatively small. We took 4 nm, which it is a typical NP-NP distance in the DNA-origami assemblies [29]. (2) The idea behind the design of the Au-Ag Fano dimer in Fig. 3 is to use two different materials. The Au NP should be large and should have a broad plasmonic peak (see Fig. S3b). Simultaneously, the other element has to have a very narrow plasmonic peak. A plasmonic NP that satisfies these requirements is a small Ag nanorod (see Fig. S3c). Its plasmonic resonance can be tuned to match the center of the broad plasmonic peak of the Au NP. This Ag nanorod should be normal to the surface of the Au sphere because this configuration leads to the strongest NP-NP interaction in this hybrid dimer. Again, the NP-NP distance should be taken small enough to have a strong coupling. Using the above ideas and principles, we can design a dimer with the spectrum exhibiting a strong and prominent Fano effect (see Fig. 3b). To finish this consideration, one should mention about the possibility to tune the NP-NP gap. The Supporting Information shows such data. In Fig. S11, we vary the NP-NP gap in the case of the Drude dimer made of two Drude metals. By increasing the gap, one can see a smooth transition from the Rabi regime to the Fano one. As the system undergoes the Rabi-to-Fano transition, the number of beats in the temporal dynamics drops from two to zero, with the typical slowly decaying tail at long times previously described (see Fig. S12).

Theoretically, only a few studies have been published so far on the coherent time dynamics of surface plasmons in NP assemblies [30,31,29,32]. Technologically, Fano and Rabi dimers can be fabricated with lithographic techniques [33-35] or using bio-assembly [29]. One powerful tool to observe single NP responses in the frequency domain is dark-field spectroscopy, i.e. the measurement of scattering in the far field. This method was successfully applied to a variety of plasmonic NPs [13, 36-39]. Moreover, this method can be also used in time-resolved studies [39]. In our study, we predict that the Fano dimer should show a unique relaxation pattern in the dynamics of the induced dipole (Figs. 1f, 3b, S8 and S9), a behavior that can be observed with time-resolved far-field scattering. The point-dipole excitation, used in our study to demonstrate the time dynamics and the related NP-NP energy transfer, can be realized experimentally with plasmonic tip-based spectroscopies, such as nanoscopy [40] or SNOM [41]. The time resolution, which is needed to observe the predicted Fano dynamics, is in the range of 5-10 fs. Simultaneously, the majority of the current experiments in the field of ultrafast spectroscopy have been performed with the time resolution of ~80 fs [13, 42-44], which is not sufficient for our effects. However, the area of time-resolved spectroscopies develops fast and experiments with a very high resolution, in the range of ~ 1-15 fs already exist [39, 45-48]. Along with the all-optical approaches cited above, one should mention a proposal on sub-fs electron-beam microscopy [32] that can also be applied to observe the coherent plasmonic dynamics described in our study.

Regarding the potential outlook for the predicted effects, the field of nanostructures is unique since it allows us to realize in practice physical models of our choice. As mentioned above, potential experiments on the Fano dynamics of plasmonic excitations can be performed either with lithographic samples or with DNA-assembled NP complexes. The latter case offers more precise control over the interparticle gaps [29]. However, the lithographic systems can be more uniform and, therefore, would not require single-particle measurements. In our plasmonic dimers, the Fano and Rabi effects come from the interference of localized excitations. Furthermore, the Fano-like effects and their time-dependent manifestations can also appear in purely photonic systems, which couple strongly with far fields [49]. Another important point we need to make is that our predictions relate to the plasmonic systems described by classical electrodynamics. In other words, our models are classical. The Fano and Rabi effects in semiconductor quantum dots and other quantum systems require very different treatments, which are based on quantum mechanics. Therefore, such systems should be studied separately, and our classical results may not apply to them.

# 4   Conclusions

We have studied the time-domain dynamics of plasmonic dimers, with the emphasis on the Fano and Rabi regimes. Whereas single NPs show trivial time dynamics and fast decays with no coherent energy transfer (i.e. no beats), both kinds of dimers show nontrivial decay dynamics with notable features due to coherent energy transfer. The fundamental difference between the time dynamics of the Rabi and Fano systems is the number of beats. The Rabi system may have any number of beats, whereas the Fano system may have at most one. Another very characteristic feature of the Fano dimer is a prominent long-lasting tail in its response. Moreover, the peculiar dynamical properties of the strongly interacting dimers are related to the spectral properties of the isolated NPs constituting the dimers. Our results could be tested with already available structures and the rapidly developing field of ultra-fast spectroscopies. As such, we expect that our results will motivate further research within temporal coherent plasmonics.

# 5   Supplemental Material

The supplementary material is available online on the journal's website or from the author

# 6   Acknowledgments

O.A.-O. and A.O.G were supported by the Nanoscale & Quantum Phenomena Institute at Ohio University. L.V.B was supported by the Institute of Fundamental and Frontier Sciences, University of Electronic Science and Technology of China and China Postdoctoral Science Foundation (2017M622992 and 2019T120820). Z.M.W. was funded by the National Key Research and Development Program (No. 2019YFB2203400), the "111 Project" (B20030) and the UESTC Shared Research Facilities of Electromagnetic Wave and Matter Interaction (Y0301901290100201).

# Supplementary information

# Temporal Plasmonics: Fano and Rabi regimes in the time domain in metal nanostructures

**Oscar Ávalos-Ovando, Lucas V. Besteiro, Zhiming Wang, and Alexander O. Govorov**

### I. Theoretical Modeling

#### 1. Modeling of nano particle systems

Systems composed of metal nanoparticles immersed in a dielectric matrix were simulated within a classical electrodynamical approach with Comsol Multiphysics, by using standard boundary conditions. We treated two kinds of metal particles, in regards to their material composition: Drude-like dielectric constants, and realistic systems (gold and silver). In all the calculations we used the optical dielectric constant of the matrix as that of water, given by $\varepsilon_w$=1.8. Absorption, scattering and extinction optical cross sections were simulated with light polarized along the z-axis, and approaching the NP along the x-axis. In the case of nanorods and dimers, the rod's (dimer's) axis was laid along the z-axis.

#### 2. Classical simulations

Realistic Au and Ag systems were simulated using the dielectric constants published by Johnson and Christy [S1], whereas the Drude-like dielectric constants can be tailored to highlight the relevant physics. Our two Drude models used in Fig. 1 in the main text were chosen in order to simulate optical profiles with two different kinds of plasmonic resonances, a sharp one and a broad one, so that the interplay between them will lead to Fano effect (Fig. S1b), and the interplay of two sharp resonances will lead to a Rabi regime. The relative permittivity within the Drude model is given by

$$\varepsilon_i = \varepsilon_{i,b} - \frac{\omega_{i,p}^2}{\omega \cdot (\omega - i\gamma_i)},$$

where $\varepsilon_{i,b}$ is the long-wavelength background dielectric constant, $\omega_{i,p}$ is the plasma frequency, and $\gamma_i$ is the damping coefficient [S2], for NPs $i = NP_1, NP_2$. In order to simulate the two resonances described before, we used two different values of $\gamma_i$ and obtained two plasmonic oscillators, one with a very sharp resonance and another one with a very broad line (Fig. 1e). The parameters of the Drude models were chosen to obtain the plasmonic resonances in the visible. Also, in this expression, $\omega$ is given in units of eV, as $\omega = 2\pi\hbar c/\lambda$. The following table summarizes the Drude parameters:

| Model | $\omega_{i,p}$ | $\gamma_i$ | $\varepsilon_{i,b}$ |
|---|---|---|---|
| Drude 1 | 12 eV | 0.001 eV | 18 |
| Drude 2 | 12 eV | 0.6 eV | 18 |

Details of these dielectric functions are shown in Figure S1a.

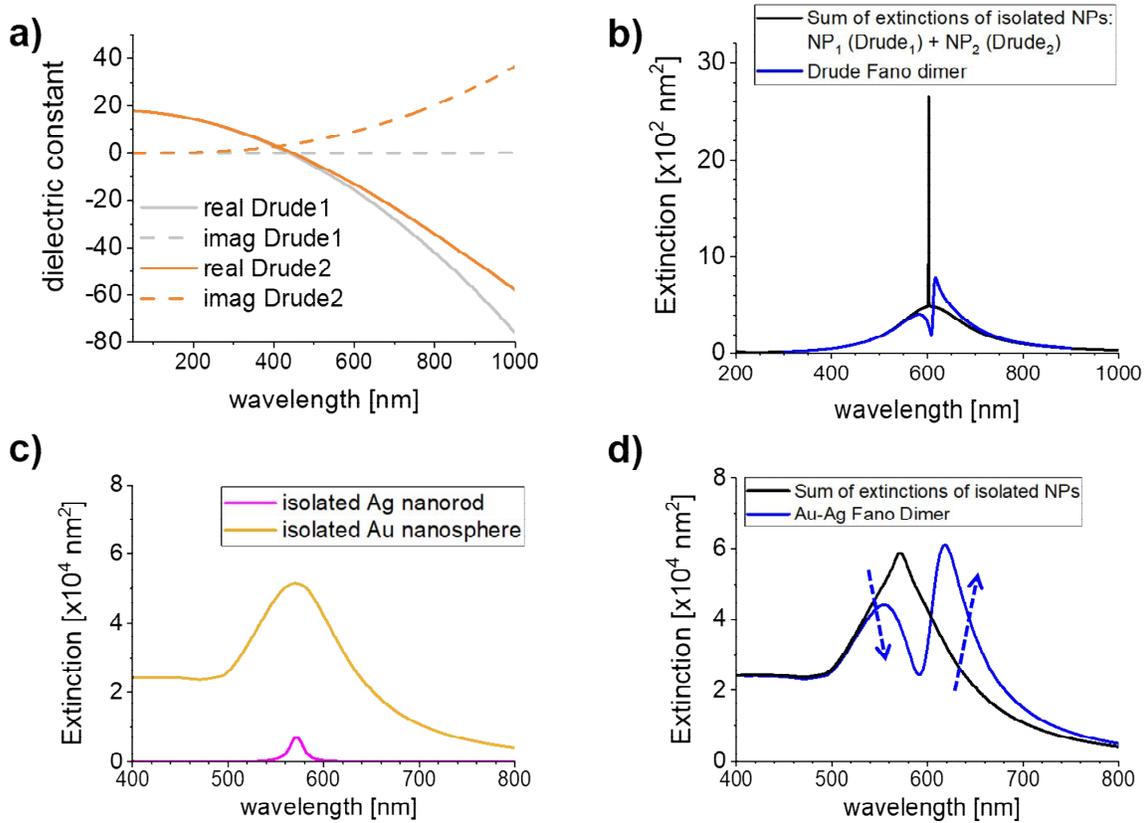

**Figure S1: (a,b)** Properties of the Drude systems. Panel (a) shows the dielectric constants. In the panel (b), we show the creation of the Fano effect in the Fano dimer of Fig. 1(d-f) in the main text. **(c,d)** Extinction cross sections related to the Au-Ag Fano dimer of Fig. 3 in the main text. Panel (c) shows the extinctions of the single non-interacting NPs, whereas the panel (d) includes the sum of the previous extinctions and the extinction of the Au-Ag complex. The panel (d) shows the appearance of the Fano effect, which comes from the interference inside the Au-Ag dimer. The extinction increases on the red side of the resonance (constructive interference) and, simultaneously, the signal decreases on the blue slope of the main resonance (destructive interference), as indicated by the dashed blue arrows.

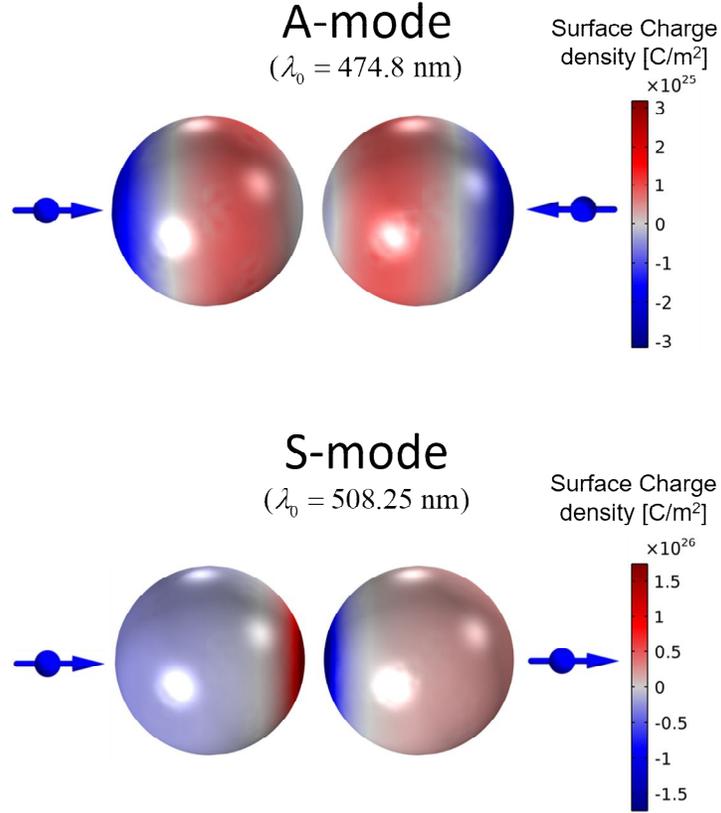

**Figure S2:** Anti-symmetric and symmetric plasmons from the Rabi dimer of Fig. 1(a-c) in the main text, for wavelengths 474.8 nm and 508.2 nm, respectively.

### 3. Time dynamics formalism

**3.1 General equations**. Time dynamics follow by the most general form of the Maxwell's equations:

$$\nabla \cdot \mathbf{D} = \rho$$
$$\nabla \times \mathbf{H} - \frac{\partial \mathbf{D}}{\partial t} = \mathbf{j}$$
$$\nabla \times \mathbf{E} + \frac{\partial \mathbf{B}}{\partial t} = 0 \tag{S1}$$
$$\nabla \cdot \mathbf{B} = 0$$

where $\mathbf{j}$ and $\rho$ are the external current and charge densities, respectively. Then, we note that the displacement field and the electric field are related via the equality, which incorporates the Fourier integral:

$$\mathbf{D}(\mathbf{r},t) = \int_{-\infty}^{\infty} e^{+i\omega \cdot t} \mathbf{D}_\omega(\mathbf{r}) \cdot d\omega = \int_{-\infty}^{\infty} e^{+i\omega \cdot t} \varepsilon_0 \varepsilon_\omega(\mathbf{r}) \mathbf{E}_\omega(\mathbf{r}) \cdot d\omega \,. \quad (S2)$$

For the magnetizing field $\mathbf{H}$ and the magnetic field $\mathbf{B}$ in non-magnetic materials, we have:

$\mathbf{H} = \dfrac{1}{\mu_0} \mathbf{B}$. Because of the causality principle, the dielectric function in the frequency domain should obey the symmetry relations:

$$\operatorname{Re}[\varepsilon_\omega(\omega)] = \operatorname{Re}[\varepsilon_\omega(-\omega)]$$
$$\operatorname{Im}[\varepsilon_\omega(\omega)] = -\operatorname{Im}[\varepsilon_\omega(-\omega)]$$

The external current in (S1) has the form of the delta function

$$\mathbf{j}(\mathbf{r},t) = j_0(t) \cdot \hat{\mathbf{z}} \cdot \delta(\mathbf{r} - \mathbf{r}_d) \,.$$

In our calculations, this current comes from the exciting point dipole:

$$\mathbf{d}(\mathbf{r},t) = d_0(t) \cdot \hat{\mathbf{z}} \cdot \delta(\mathbf{r} - \mathbf{r}_d) \,,$$

where the point dipole is taken to oscillate along the z-direction (i.e. parallel to the dimer axis) in order to excite the strongest longitudinal modes in our system; $\mathbf{r}_d$ is the dipole position and $d_0(t)$ is the time-dependent dipole strength. The relations between the current flux, the charge density and the dipole are coming from the continuity equation, which yields for us:

$$\rho(\mathbf{r},t) = A_0(t) \cdot \delta'_z(\mathbf{r} - \mathbf{r}_d)$$
$$\frac{\partial \rho(\mathbf{r},t)}{\partial t} = -j_0(t) \cdot \delta'_z(\mathbf{r} - \mathbf{r}_d), \quad \frac{\partial d_0(t)}{\partial t} = j_0(t) \quad (S3)$$

In the following steps, the numerical problem of the excitation will be solved using direct and inverse Fourier transforms:

$$d(\mathbf{r},t) = \int_{-\infty}^{\infty} e^{+i\omega \cdot t} d_\omega(\mathbf{r}) \cdot d\omega, \quad d_\omega(\mathbf{r}) = \frac{1}{2\pi} \int_{-\infty}^{\infty} e^{-i\omega \cdot t} d(\mathbf{r},t) \cdot d\omega,$$

Then, the Maxwell's equations should be Fourier-transformed, and we obtain:

$$\nabla \cdot \varepsilon_0 \varepsilon_\omega(r) \mathbf{E}_\omega = \rho_\omega$$
$$\nabla \times \mathbf{B} - \mu_0 i\omega \mathbf{D} = \mu_0 \mathbf{j}_w , \text{ (S4)}$$
$$\nabla \times \mathbf{E} + i\omega \mathbf{B} = 0$$
$$\nabla \cdot \mathbf{B} = 0$$

The external current here should be taken in the Fourier space:

$$\mathbf{j}_\omega = i\omega \cdot d_{0,\omega} \cdot \hat{\mathbf{z}} \cdot \delta(\mathbf{r} - \mathbf{r}_d) .$$

**3.2 Green function formalism.** Now we introduce the Green function for the point dipole. This function will be denoted as $G_d(\omega)$ and it is the response to the external point dipole

$$\tilde{\mathbf{d}}(\mathbf{r},t) = e^{i\omega t} \cdot \hat{\mathbf{z}} \cdot \delta(\mathbf{r} - \mathbf{r}_d) ,$$

The corresponding current density follows from (S3):

$$\tilde{\mathbf{j}}(\mathbf{r},t) = i\omega \cdot \hat{\mathbf{z}} \cdot \delta(\mathbf{r} - \mathbf{r}_d) \cdot e^{+i\omega \cdot t}, \quad\quad\quad \text{(S5)}$$

For all induced quantities related directly to the Green function, we will adopt the notations like $\tilde{\mathbf{d}}$, $\tilde{\mathbf{j}}$, etc. Then, the Green function $G_d(\omega)$ should be calculated as the dipole moment induced on the NP by the perturbation (S5):

$$\tilde{d}_{NP}(t) = \int_{-\infty}^{\infty} \tilde{\rho}_{NP}(\mathbf{r},t) \cdot x \cdot dV = \tilde{d}_{0,NP} \cdot e^{+i\omega \cdot t}$$
$$G_d(\omega) = \tilde{d}_{0,NP}$$

where $\tilde{\rho}_{NP}(\mathbf{r},t) = \tilde{\rho}_{0,NP}(\mathbf{r}) \cdot e^{+i\omega \cdot t}$ is the induced surface change density.

Now we consider a Gaussian excitation pulse and look at the induced dipole moment on a NC:

$$d_{NP}(t) = \int_V \rho_{NP}(\mathbf{r},t) \cdot z \cdot dV = \int_S \sigma_{NP}(\mathbf{r},t) \cdot z \cdot dS ,$$

where $\sigma_{NP}(\mathbf{r}, t)$ is the surface charge on a NP. The response of a NP to the pulsed excitations should be written as

$$d_{NP}(t) = \int\limits_{-\infty}^{\infty} e^{+i\omega t} d_0(\omega) \cdot G_d(\omega) \cdot d\omega,$$

where $d_0(\omega)$ is the Fourier transform from the Gaussian pulse. The mathematical expressions for the pulse and for its Fourier transform are here:

$$d_0(t) = d_0 e^{+i\omega_0 t - \left(\frac{t}{\Delta t}\right)^2}, \quad d_0(\omega) = \frac{d_0}{2\sqrt{\pi}} \Delta t \cdot e^{-\frac{1}{4}\Delta t^2 (\omega - \omega_0)^2}.$$

Another type of excitation involves a pulsed plane wave coming from one direction, e.g. along the +x direction. In this case, the excitation current density is zero, i.e. $\mathbf{j}_\omega = 0$, but the system experiences an external field:

$$\mathbf{E}_{ext}(t, \mathbf{r}) = \hat{\mathbf{z}} \cdot E_0 \cdot e^{i\omega_0(t - x/c) - \left(\frac{t - x/c}{\Delta t}\right)^2} = \hat{\mathbf{z}} \cdot E_0(x, t).$$

The corresponding solution of the problem for the induced dipole in terms of the Green function is now given by:

$$d_{NP}(t) = \int\limits_{-\infty}^{\infty} e^{+i\omega t} E_{0,\omega} \cdot G_E(\omega) \cdot d\omega, \qquad E_{0,\omega} = E_{0,\omega}(x = 0).$$

where $G_E(\omega)$ is the corresponding Green function and it is the response to the external field in the form: $\tilde{\mathbf{E}}_{ext}(t, \mathbf{r}) = \hat{\mathbf{z}} \cdot e^{i\omega(t - x/c)}$; the coordinate x=0 is the center of a dimer. The solution for the electric field in our system would be also given via a Green function:

$$\mathbf{E}(\mathbf{r}, t) = \int\limits_{-\infty}^{\infty} e^{+i\omega t} E_{0,\omega} \cdot \mathbf{F}_E(\mathbf{r}, \omega) \cdot d\omega,$$

where $\mathbf{F}_E(\mathbf{r}, \omega)$ is the corresponding Green function for the electric field and it is a vector.

## II.Real systems

### 1. Single nanoparticles

For the dimers described in the main text, we used a spherical Ag NP, a spherical Au NP, and a Ag nanorod. The optical properties and time dynamics of these single NPs are shown in Fig. S3, in the right and left panels, respectively. The optical cross sections of all single NPs show a single plasmonic resonance at different optical frequencies $\omega_0$, with FHMWs of 146 meV, 492 meV, and 55 meV, for panels (a-c) respectively. Around each of these resonances, we set up a Gaussian excitation pulse (blue solid line) with the right duration $\Delta t$ in order to capture the plasmon and to avoid interband transitions. Then, each NP is illuminated with a pulsed electric field tailored with $\omega_0$ and $\Delta t$, from which we calculated the total dipole moment and extracted its inverse Fourier transform, shown in the right panels of Fig. S3. These time dynamics are characterized by a single envelope exponential decay $\sim \exp\left[-t/\tau\right]$ (rather than a number of beatings as in the case of dimers), with $\tau_{AgNP}$=8.98 fs, $\tau_{AuNP}$=2.71 fs, and $\tau_{AgNR}$=25 fs, for panels (a-c) respectively. These values are in excellent agreement with estimations from the FWHMs of the extinction coefficients, in where $\tau \approx 2\hbar/\text{FWHM}$, which leads to $\tau_{AgNP} \approx$9.02 fs, $\tau_{AuNP} \approx$2.7 fs, and $\tau_{AgNR} \approx$24 fs, for panels (a-c) respectively.

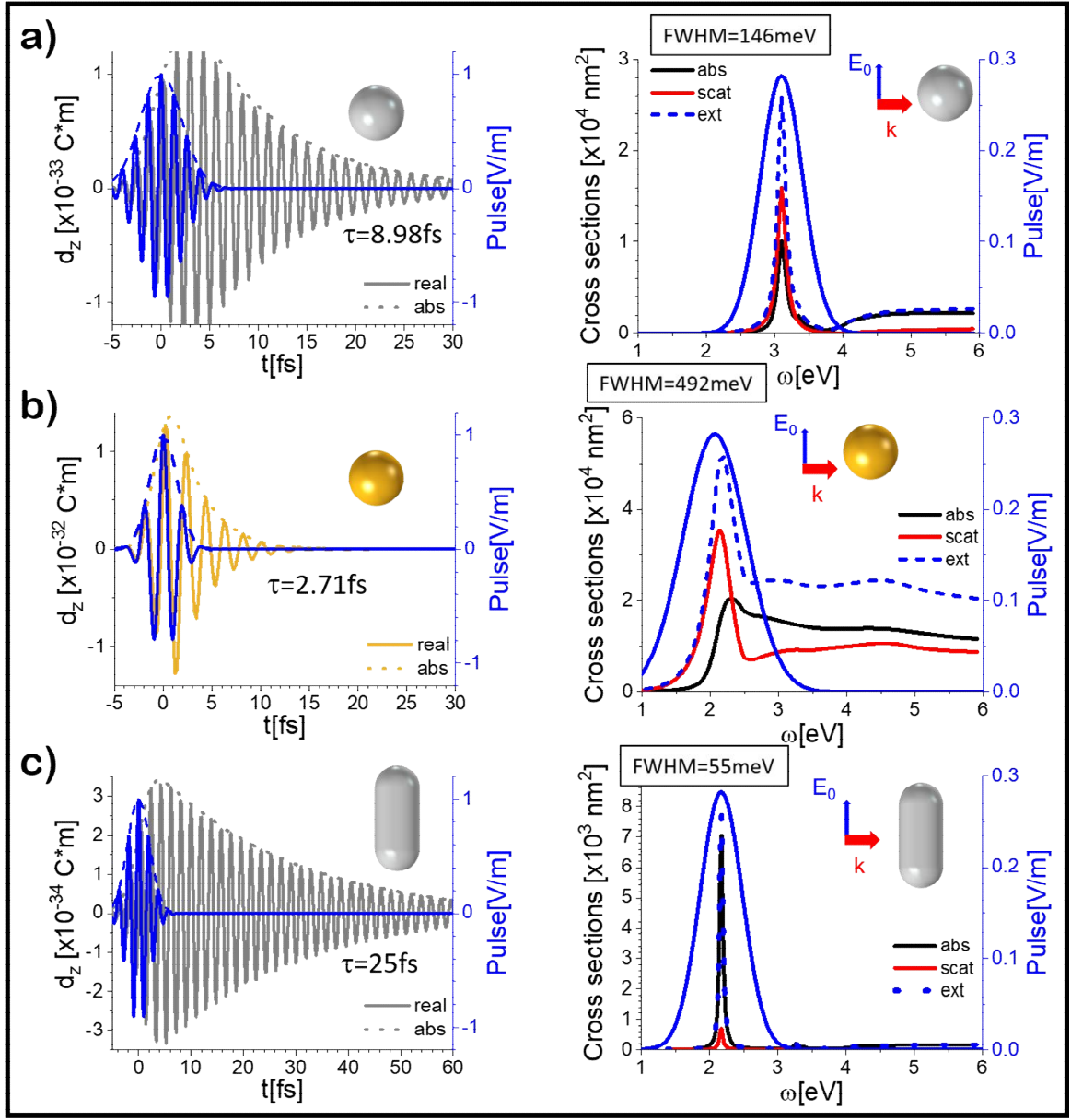

**Figure S3:** Dynamic properties of single NPs. We consider here the following NPs: Ag sphere (a), Au sphere (b), and Ag nanorod (c). The sizes of the NPs in this figure are identical to the ones used in the main text. Left panels: the time dynamics of the total dipole moment of a NP and the exciting pulses. Right panels: the optical cross sections, the pulse spectra, and the schematics of incident light. Here we consider an incident wave polarized along the z-direction with $E_0$=1 V/m (blue arrow). In (a), the parameters are the following: $R_{Ag\text{-}NP}$=20

nm, Δt=3 fs, and **ω₀**=3.1 eV. In (b), we use: $R_{Au-NP}$=50 nm, Δt=2 fs, and **ω₀**=2.07 eV. In (c), we choose: $R_{Ag-NR}$=5 nm, $L_{Ag-NR}$=27 nm, Δt=3 fs, and **ωₙ**=2.17 eV.

### III. Time dynamics in details.

Here we give detailed information for the time traces for the main subsystems of our interest. We note that, again, the Rabi and Fano regimes show remarkably different time dynamics, and that the main features of the Fano regime also occur when the excitation for the Fano Au-Ag dimer is done with the pulsed plane electromagnetic wave (Fig. S8). In addition, we observe that the Au-Au dimer does not show the Rabi oscillations, and this is because of the strong dissipation in Au (Fig. S7).

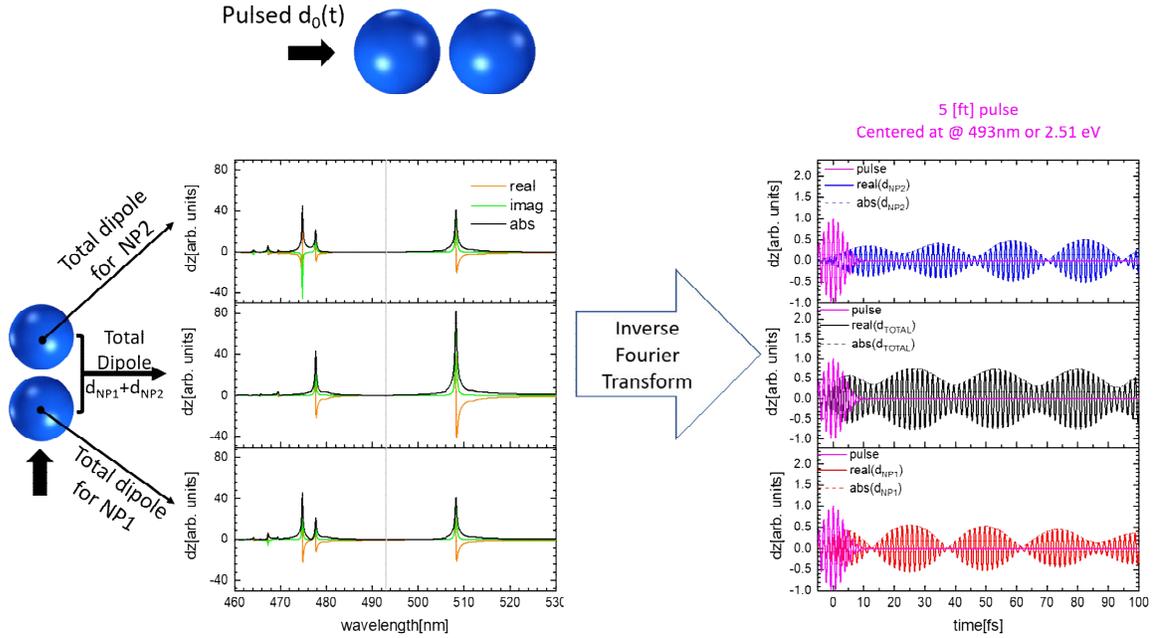

**Figure S4:** Excitation with a point dipole (black arrow) of Rabi Drude dimer with $d_0(t)$ as in Figs.1a-c of main text.

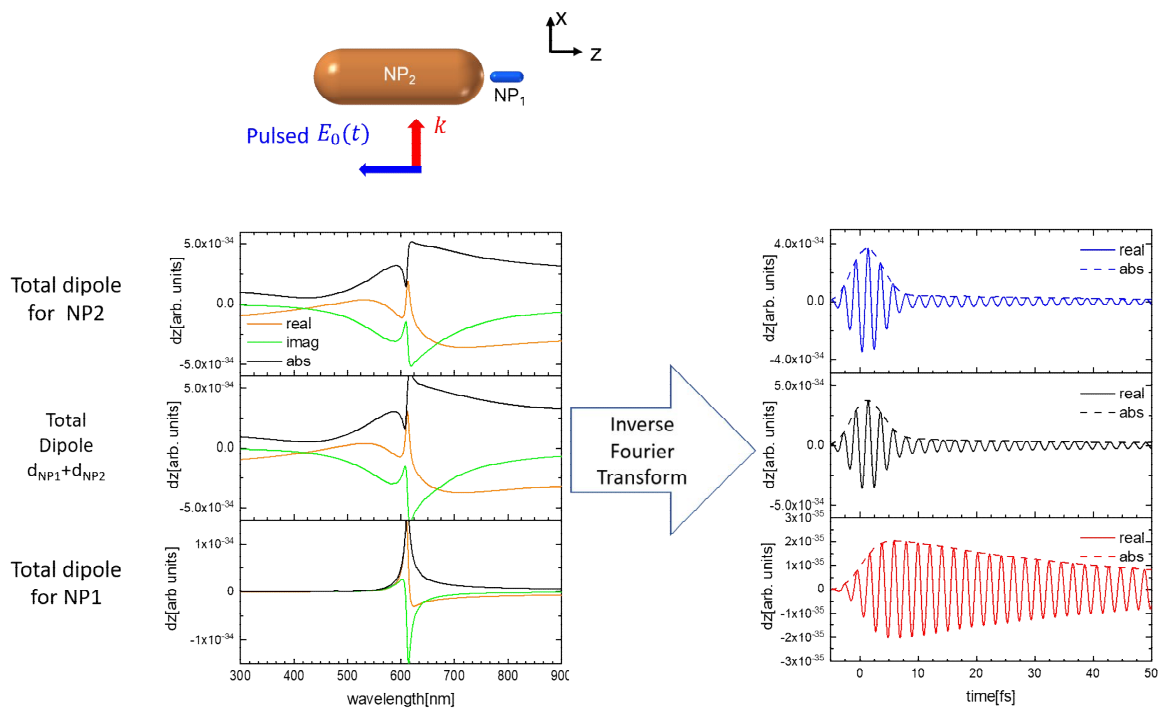

**Figure S5:** Excitation of Fano Drude dimer with $E_0(t)$ as in Figs.1d-f of main text.

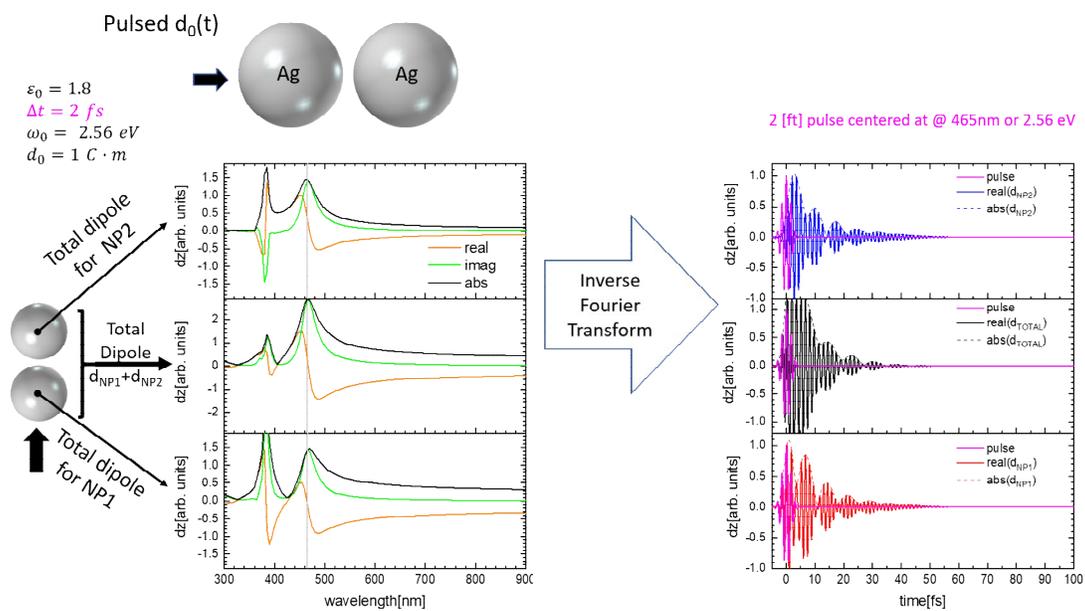

**Figure S6:** Excitation of Rabi Ag-Ag dimer with $d_0(t)$ in details.

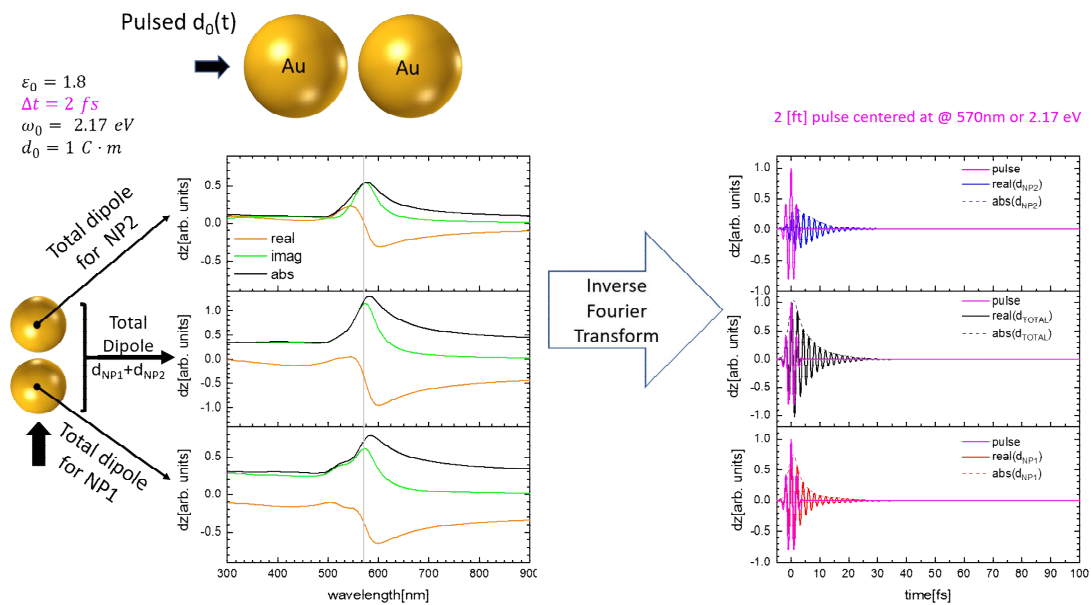

**Figure S7:** Excitation of Rabi Au-Au dimer with $d_0(t)$ in details.

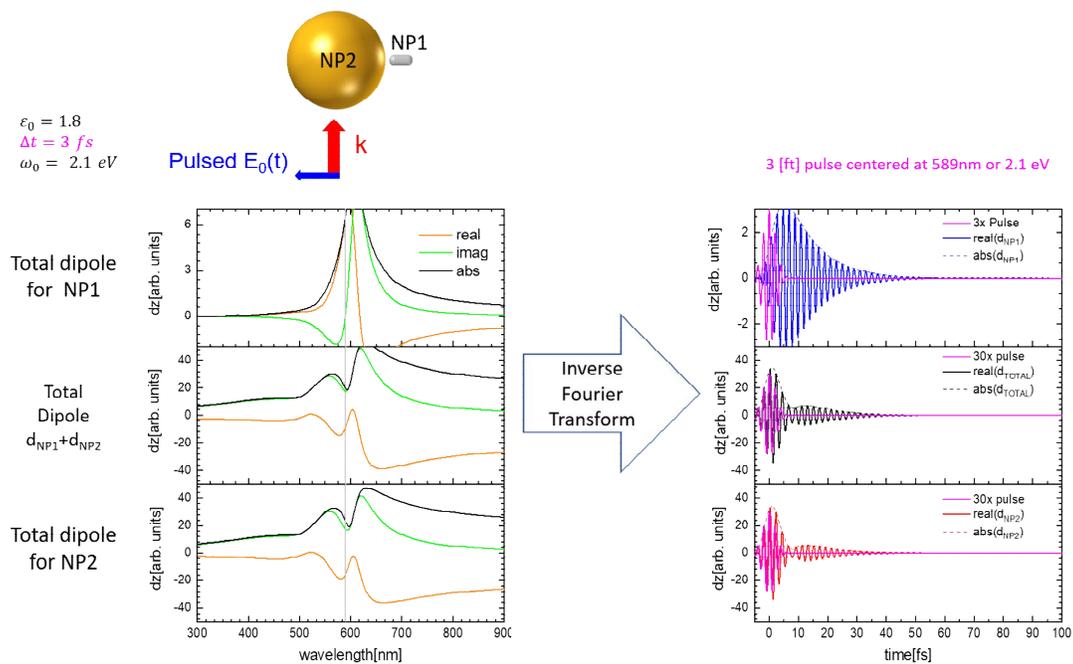

**Figure S8:** Excitation of Fano Au-Ag dimer with $E_0(t)$ in details.

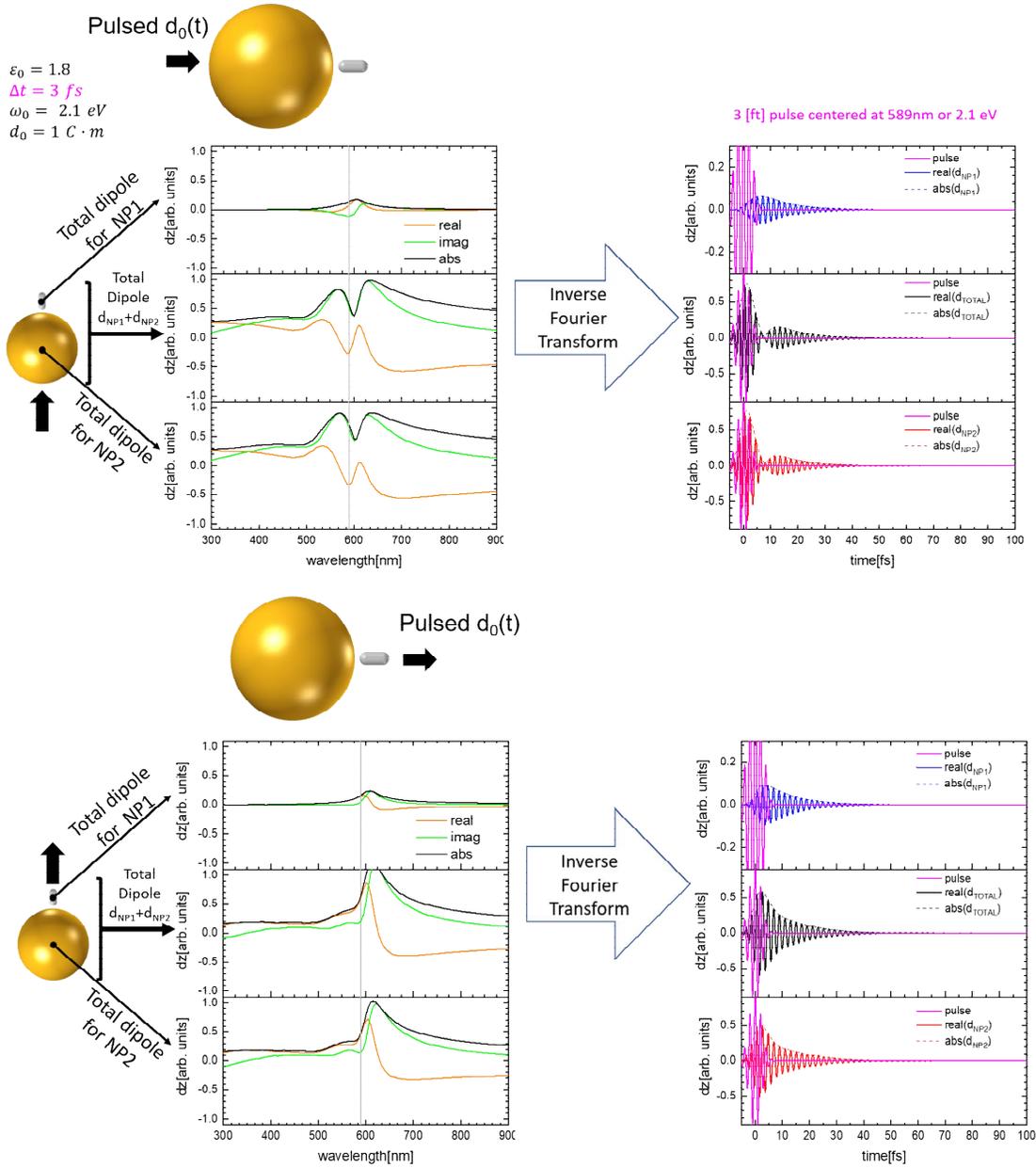

**Figure S9:** Excitation of Fano Au-Ag dimer with $d_0(t)$ in details. Here we show two geometries of excitation with the point dipole (black arrow). We observe that the dynamics for the total induced dipole in the dimer are different. When the dimer is excited with the point dipole on the right (bottom panels), the whole system does not show beats in the dynamics of $d_{tot}(t)$.

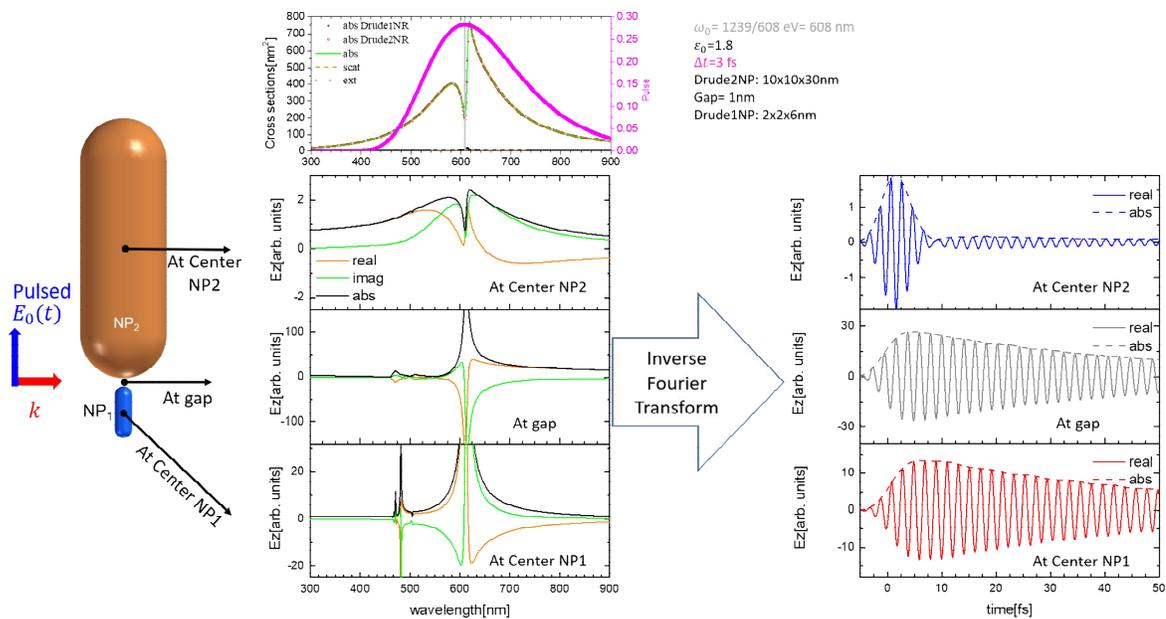

**Figure S10:** Non-local responses in the Fano dimer made of Drude metals. Excitation is with $E_0(t)$. We observe here that the character of coherent dynamics depends on the spatial position.

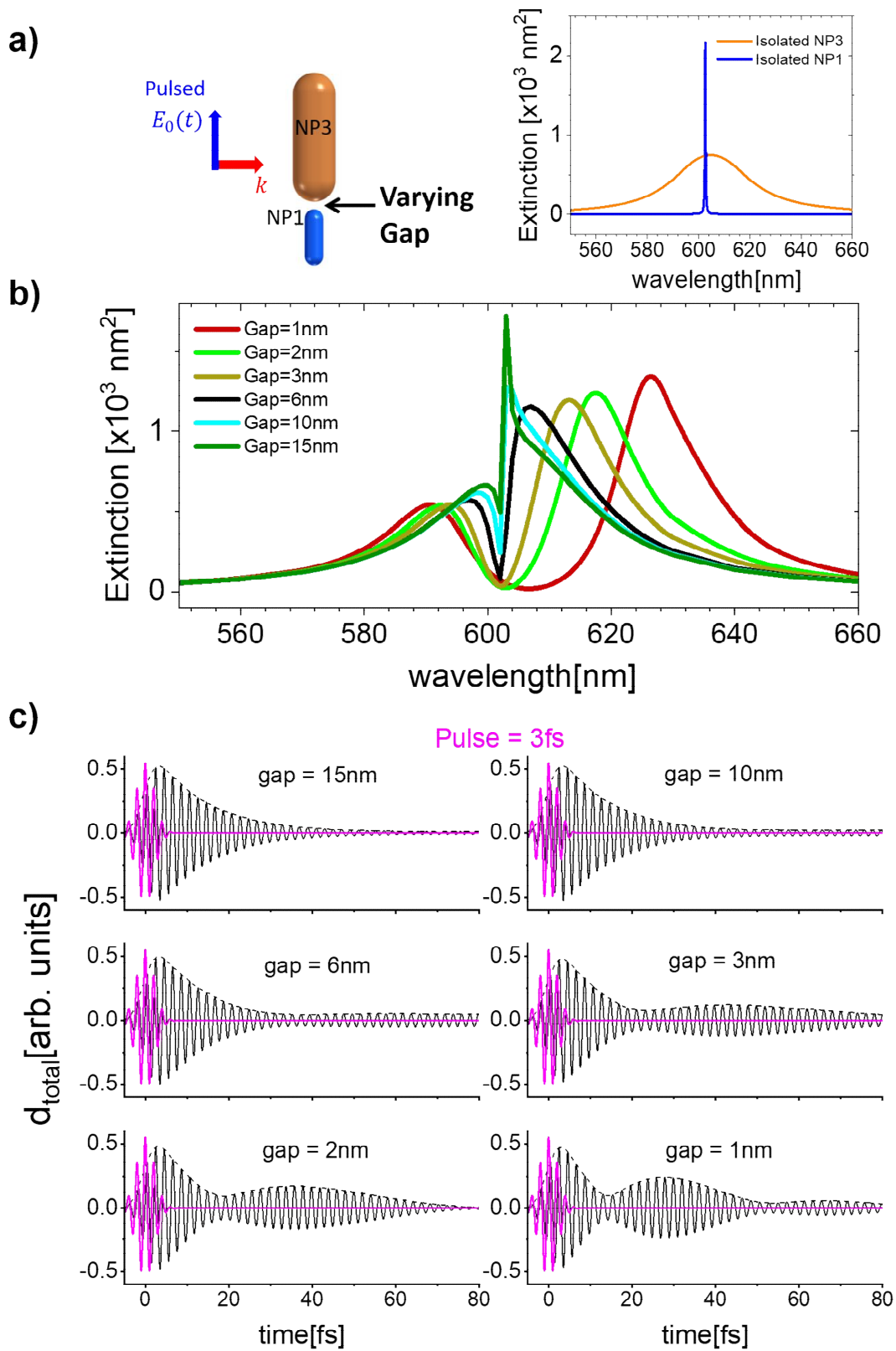

**Figure S11:** Computational data for the Fano dimer based on the Drude metals (panel a). Here we vary the NP-NP gap and observe a smooth transition between the Fano and Rabi regimes (panel b). The number of the beats changes from zero to two (panel c). The parameters for the NP1 are the same as in Fig. 1d of the main text ($R_{NP1}$=1 nm, $L_{NP1}$=6 nm). For the NP3, we use the following parameters: $\omega_{3,p} = 12$ eV, $\varepsilon_{3,b} = 18$, $\gamma_3$ =0.1 eV, $R_{NP3}$=3.3 nm, and $L_{NP3}$=20 nm. We also assume that $\Delta$t=3 fs and $\lambda_0$=604 nm, where $\lambda_0 = 2\pi c / \omega_0$.

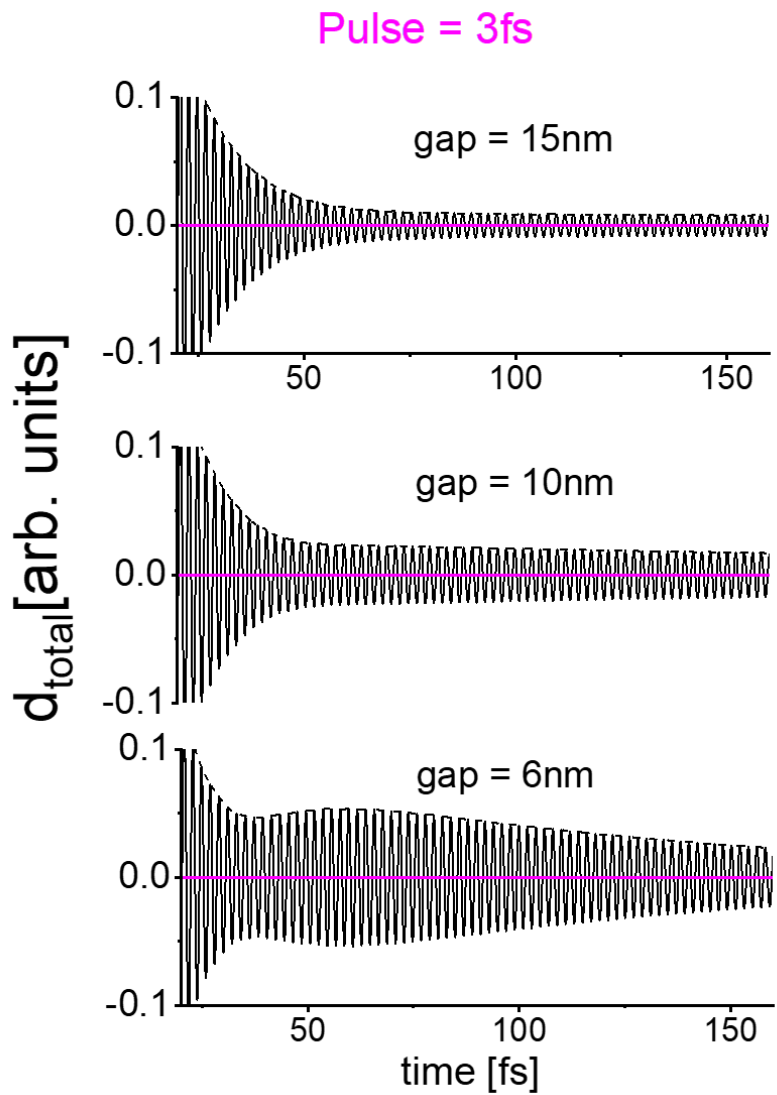

**Figure S12:** The same data as in Figure S11 but focusing on long times. We show the appearance of the long-lasting tails for the cases: gap=15nm, 10nm, and 6nm. In the cases with gap=15nm and 10nm, the tail decreases monotonically, like for the Fano dimer in Fig. 1f in the main text. For the Fano dimer with gap=6nm, the NP-NP interaction becomes strong enough, and now the time dynamics show an additional oscillation.

**References (for Supplemental Information)**